\documentclass[10pt,twocolumn,superscriptaddress,nofootinbib]{revtex4-1}

\usepackage{natbib}
\usepackage{textcase}
\usepackage{bm}
\usepackage{graphicx}
\usepackage{url}
\usepackage{braket}
\usepackage{amsmath,amssymb}  
\usepackage{subfigure}
\usepackage{nicefrac}
\allowdisplaybreaks

\def\i{\mathrm{i}}
\def\s{\mathrm{s}}
\def\p{\mathrm{p}}

\usepackage{color}
\usepackage{layouts}
\graphicspath{{Figures/}{Pictures/}{PSTricks/}} 

\begin{document}


\title{Engineering spectrally unentangled photon pairs from nonlinear microring resonators through pump manipulation}


\author{J.~B.~Christensen}
\email[]{jesbch@fotonik.dtu.dk}
\affiliation{Department of Photonics Engineering, Technical University of Denmark, 2800 Kongens
Lyngby, Denmark.}

\author{J.~G.~Koefoed}
\affiliation{Department of Photonics Engineering, Technical University of Denmark, 2800 Kongens
Lyngby, Denmark.}

\author{K.~Rottwitt}
\affiliation{Department of Photonics Engineering, Technical University of Denmark, 2800 Kongens
Lyngby, Denmark.}

\author{C.~J.~McKinstrie}
\altaffiliation{Main address: Huawei Technologies, 400 Crossing Boulevard, Bridgewater, New Jersey 08807, USA}
\affiliation{Department of Photonics Engineering, Technical University of Denmark, 2800 Kongens
Lyngby, Denmark.}



\begin{abstract}
The future of integrated quantum photonics relies heavily on the ability to engineer refined methods for preparing the quantum states needed to implement various quantum protocols. An important example of such states are quantum-correlated photon pairs, which can be efficiently generated using spontaneous nonlinear processes in integrated microring-resonator structures. In this work, we propose a method for generating spectrally unentangled photon pairs from a standard microring resonator. The method utilizes interference between a primary and a delayed secondary pump pulse to effectively increase the pump spectral width inside the cavity. This enables on-chip generation of heralded single photons with state purities in excess of $99~\%$ without spectral filtering.   
\end{abstract}


\maketitle


\textit{Introduction}.---The integration of photonics functionalities on chip renders possible a wide range of potential applications such as signal routing \cite{Lee08}, small-scale low-threshold lasers \cite{chen2016electrically}, optical interconnects \cite{miller2009} and sensing \cite{DeVos07}. A vital photonic-chip component is the microring resonator (MRR), which is a photonic waveguide folded onto itself and sidecoupled to one or multiple accompanying bus waveguides. MRRs have been demonstrated to enable drop-filtering capabilities, biochemical-sensing, and frequency-comb generation, and have been fabricated on a wide variety of material platforms including silicon \cite{bogaerts2012silicon}, lithium niobate \cite{guarino2007electro}, aluminum gallium arsenide \cite{Pu:16}, and silicon nitride \cite{gondarenko2009high}.

In recent years, MRRs have also become a hot research topic in the context of integrated quantum photonics \cite{Politi:09,silverstone:15,harris:14,Dutt:15,Reimer:14,Preble:15,Grassani:15,vernon2015,guo:16,silverstone16,Roztocki:17}. Material platforms such as silicon exhibit a strong third-order nonlinearity, which, together with the resonantly enhanced fields inside the MRR, offer a scaleable and energy-efficient method for generating quantum-correlated photon pairs through spontaneous four-wave mixing (SFWM) \cite{Clemmen:09}. In SFWM, two pump photons at a resonance of the microring are converted into a signal and an idler photon in symmetrically surrounding resonances, obeying energy conservation. Interestingly, the photon pair can be coherently distributed over multiple resonances, which has stimulated work on quantum frequency combs for high-dimensional quantum information processing in the frequency-mode basis \cite{Reimer16}. Multiple groups have recently demonstrated coherent processing of such high-dimensional frequency-bin entangled quantum states \cite{kues2017,Imany17}.

Photon pairs originating from a single pair of microring resonances may exhibit either weak or strong spectral correlation depending on the quality factor of the resonator and the duration of the pump pulse used for excitation \cite{Helt:10}. It is often desired to generate the photon pairs such that there is no spectral correlation between paired resonances. This feature enables both high-visibility information processing of quantum frequency combs and facilitates quantum interference between single photons heralded from different photon-pair sources. Notably, such quantum interference has recently been demonstrated using heralded single photons from two silicon MRRs \cite{Imad:17,Faruque:17}. However, even with a broadband pump, the visibility of quantum interference between heralded photons from different MRRs is currently bounded to values below $\sim 92~\%$ without narrow spectral filtering \cite{Helt:10}. Such visibility values ultimately prevents up-scaling of quantum optics experiments and applications based on quantum interference. A recent proposal based on interferometrically coupling the MRR \cite{Tison17}, permits a broadening of the pump resonance relative to the sideband resonances, nearly eliminating spectral correlations \cite{vernon:17}. However, such interferometric coupling complicates the MRR structure, and has limited use for quantum-frequency-comb generation, which requires many identical sideband resonances.

In this Letter, we propose an alternative method for generating spectrally unentangled photon pairs. Our scheme works for standard MRRs and relies only on pumping the ring with a superposition of two phase-shifted and temporally displaced Gaussian pulses. We show that this enables a broader in-resonator pump spectrum, that increases the heralded single-photon purity beyond the usual $92~\%$ bound.



\textit{Theory and model}.---Third-order nonlinear MRRs driven on resonance by a coherent pump (p) enables growth of quantum-correlated signal (s) and idler (i) fields in the surrounding resonances, with angular frequencies constrained by $2\omega_\p = \omega_\s + \omega_\i$. While pairs of so-called 'twin beams' may appear in multiple resonances symmetrically around the pump resonance, we limit ourselves to considering only a single signal-idler pair, and note that the extension to multiple resonances is possible \cite{Helt:10}. Furthermore, we consider the case where the pump is spectrally overlapping with only a single resonance, which we shall refer to as the pump resonance. Consequently, we are in the non-transient regime, in which the pump-pulse duration is much longer than the round-trip time of the MRR. 


For low pump powers, only single pairs of signal and idler photons are probabilistically generated inside the MRR. The bi-photon-part of the output quantum state is expressed as
\begin{equation}
\ket{\Psi_{\textrm{II}}} = \iint  \mathrm{d}\omega_\i \mathrm{d}\omega_\s \mathcal{A} (\omega_\i, \omega_\s )\hat{a}_\i^\dagger(\omega_\i) \hat{a}_\s^\dagger (\omega_\s) \ket{\mathrm{vac}},
\label{eq:biphotonstate}
\end{equation}
wherein $\mathcal{A}(\omega_\i, \omega_\s)$ is the bi-photon joint spectral amplitude (JSA), and $\hat{a}_j^\dagger(\omega_j)$ is a creation operator of a photon in mode $j \in \left\lbrace    \i , \s  \right\rbrace $, assumed to have no spectral overlap with adjacent resonances. 

An important property for a photon-pair source, is the degree of spectral entanglement between the signal and idler photons. Such spectral entanglement can be characterized by a Schmidt decomposition of the JSA, i.e.~\cite{Uren05}
\begin{equation}
\mathcal{A} (\omega_\i, \omega_\s ) = \sum_{k}  \lambda_k \psi_{\i,k} (\omega_\i) \psi_{\s,k} (\omega_\s) ,
\label{eq:Schmidt}
\end{equation}
where $\lambda_k$ are descendingly-ordered real-valued Schmidt coefficients, and $\left\lbrace \psi_{k} \right\rbrace $ are orthonormal sets of idler and signal wavepacket (Schmidt) modes. A large degree of spectral entanglement manifests itself by many non-zero, and comparable, Schmidt coefficients, while a complete lack of spectral entanglement entails that the JSA is factorable, $\mathcal{A} (\omega_\i, \omega_\s ) =   \lambda_1 \psi_{\i,1} (\omega_\i) \psi_{\s,1} (\omega_\s) $. Using the Schmidt decomposition, the quantum-state purity of a heralded photon is expressed by 
\begin{equation}
\mathcal{P} =\frac{1}{R^2} \sum_{k} \lambda_k ^4 = \sum_{k} \lambda_k ^4 \Big/ \Big( \sum_{m}\lambda_m ^2 \Big)^2,
\label{eq:Purity}
\end{equation}
and is unity only in the case of a factorable JSA. In Eq.~\eqref{eq:Purity}, we have explicitly given the probability of a photon-pair-generation event per pump pulse $R$, which is restricted to $R \ll 1$ to avoid the output of multi-photon states.

In the context of resonator-based sources of photon pairs using SFWM, the JSA takes the form (omitting a proportionality factor unimportant for our analysis) \cite{Helt:10}
\begin{equation}
\mathcal{A} (\omega_\i, \omega_\s ) \propto F_\p (\omega_\i + \omega_\s ) l_\i (\omega_\i) l_\s (\omega_\s ),
\label{eq:JSA}
\end{equation}
in which $F_\p$ is the convolution 
\begin{equation}
F_\p(\omega) = \int \mathrm{d}\omega'  \alpha_\p (\omega-\omega') l_\p (\omega-\omega')   \alpha_\p (\omega') l_\p (\omega') ,
\label{eq:pumpconvolution}
\end{equation}
where the Lorentzian factor $l_j (\omega) = [  \omega_{j0} /(2 Q_j) + i \omega ] ^{-1}$. Here, $\omega_{j0}$ and $Q_j$ are the center frequency and the loaded quality factor of the $j$th resonance, respectively. Moreover, $\alpha_\p$ is the spectral envelope of the pump pulse prior to coupling into the MRR, and, as for the resonance linewidths $l_j$, the argument $\omega$ represents an angular frequency relative to $\omega_{j0}$.

\begin{figure}[b]
\begin{center}
\includegraphics[trim = 0mm 0mm 0mm 0mm,clip, scale=1]{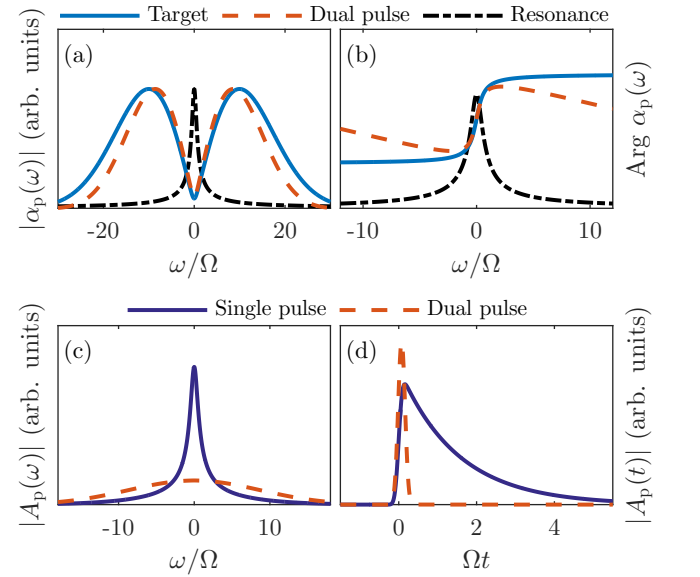}
\caption{(a) Pulse spectrum $\vert \alpha_\p(\omega) \vert $, and (b) spectral phase, of the `target' pulse and the dual-pulse configuration, both shown alongside the corresponding resonance spectrum $\vert l_\p (\omega) \vert$. (c) Spectral- and (d) temporal distribution of the in-resonator pump field for both single- and dual-pulse configurations. For parameters, see text. }
\label{fig:1}
\end{center}
\end{figure}
As seen from Eq.~\eqref{eq:JSA}, the spectral correlation in the JSA is solely contained in $F_\p(\omega_\i+\omega_\s )$.  Therefore, if $F_\p$ varies slowly in comparison to $l_{\i, \s}$, the JSA is nearly factorable. However, by virtue of Eq.~\eqref{eq:pumpconvolution}, $F_\p$ is itself spectrally limited by its corresponding Lorentzian resonance $l_\p$ prohibiting an arbitrary spectral width for a Gaussian input pulse. This imposes an upper limit of $\mathcal{P} \approx 0.92$ for a Gaussian-shaped pump, which is spectrally much broader than the corresponding resonance linewidth \cite{Helt:10}. However, consider instead an incident pump spectrum of the form $\alpha_\p^{(\mathrm{tar})}(\omega) = l_\p^{-1}(\omega) \exp (-\omega^2/2\sigma^2)$. In this case, we obtain $F_\mathrm{p} (\omega)  \propto \exp( - \omega^2/4\sigma^2 )$, which results in a factorable JSA for $\sigma  \gg \omega_\p / Q_\p \equiv \Omega$. We therefore refer to this pump spectrum as the `target' spectrum. 


\begin{figure}[h]
\begin{center}
\includegraphics[trim = 0mm 0mm 0mm 0mm,clip,scale=1]{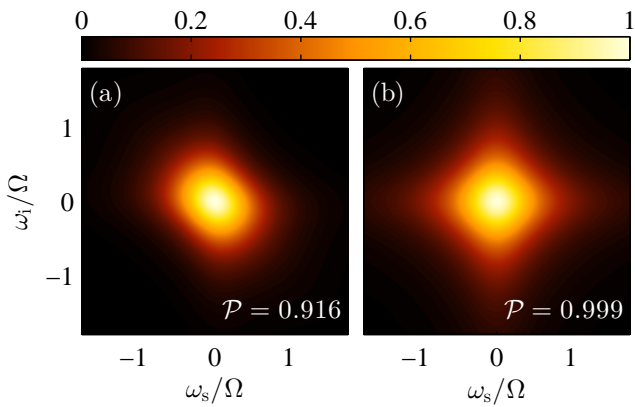}
\caption{Normalized joint spectral intensity for (a) the single-pulse case, and (b) the dual-pulse case. Axes are normalized with respect to the resonance linewidth $\Omega$. For parameters, see text.}
\label{fig:2}
\end{center}
\end{figure}

%

An approximation to the `target' spectrum is attained by superimposing two Gaussian pulses having a relative phase shift of $\pi$. Such a dual-pulse superposition is expressed by the spectral form
\begin{equation}
\alpha_\p(\omega) = \left[\sqrt{\eta} -\exp(-i  \Delta \tau \omega ) \sqrt{1-\eta } \right] \exp\left( - \tau_\p^2 \omega ^2/2 \right),
\label{eq:dualpulse}
\end{equation}
where $\tau_\p$ is the (common) pulse duration, $\Delta \tau$ is the inter-pulse temporal separation, and $\eta$ is the relative pulse weight. To see that such a dual-pulse superposition provides an excellent qualitative approximation (for certain values of $\eta$, $\Delta \tau$, and $\tau_\p$) to the target pulse spectrum, Figs.~\ref{fig:1}(a) and (b) illustrate the close resemblance, in both absolute value and phase, obtained for the parameters $\eta = 0.55$, $\tau_\p \Omega = 0.1$, $\Delta \tau \Omega=0.2 $, and $\sigma \tau_p = 1$. The spectra are significantly broader than the corresponding resonance linewidth, and they both exhibit a spectral dip coinciding with the resonance. The effect of this is seen in Fig.~\ref{fig:1}(c) which shows the in-resonator pump field $A_\p(\omega) = \alpha_\p(\omega)  l_\p(\omega)$ for the single-pulse case [$\eta = 1$ in Eq.~\eqref{eq:dualpulse}] and for the dual-pulse case parametrized as in Figs.~\ref{fig:1}(a) and (b) (here, both spectra are normalized with respect to their respective in-resonator pulse energies). Whereas the single-pulse spectrum is limited by the resonator, the dual-pulse spectrum is much broader than the corresponding resonance linewidth. This broadening effect is perhaps intuitively easier to understand from a time-domain argument. As the delayed pulse encounters the coupling region between the bus waveguide and the MRR, it interferes with the part of the early pulse still inside the resonator. Due to the relative phase shift of $\pi$ between the pulses, this interference is destructive into the resonator and constructive into the bus channel waveguide. Thus, rather than being limited by the resonator lifetime, pump light can be coupled in and subsequently out of the resonator on a time scale comparable to $\Delta \tau$. This is illustrated in Fig.~\ref{fig:1}(d) showing the stark contrast between an exponential decay and an interferometrically-induced outcoupling. The shorter lifetime effectively amounts to a smaller resonator quality factor for the pump resonance, and hence a broader pump spectrum inside the resonator, as was already shown in Fig.~\ref{fig:1}(c).

\textit{Results}.---Using Eqs.~\eqref{eq:JSA}--\eqref{eq:dualpulse}, we now demonstrate that the dual-pulse superposition can be used to eliminate the spectral correlation between the signal and idler photons. Figures \ref{fig:2}(a) and (b) illustrate normalized joint spectral intensities (JSI), given as $\vert \mathcal{A} (\omega_\i, \omega_\s) \vert ^2 $, for the single- and dual-pulse case, respectively, and using the same parameters as for Fig.~\ref{fig:1}. Even though the pump pulse is spectrally much broader than the resonance linewidth, the JSI obtained in the single-pulse case exhibits anti-correlation between the relative signal- and idler frequencies. As a result of this anti-correlation, the heralded single-photon purity, which is calculated using Eq.~\eqref{eq:Purity} and displayed in the inset, is near the asymptotic limit of 0.92 for the single-pump case. In contrast, the JSI for the dual-pulse case displays no apparent correlation between the relative signal- and idler frequencies, and shows an improvement to a near-unit purity.

\begin{figure}[t]
\begin{center}
\includegraphics[scale=1]{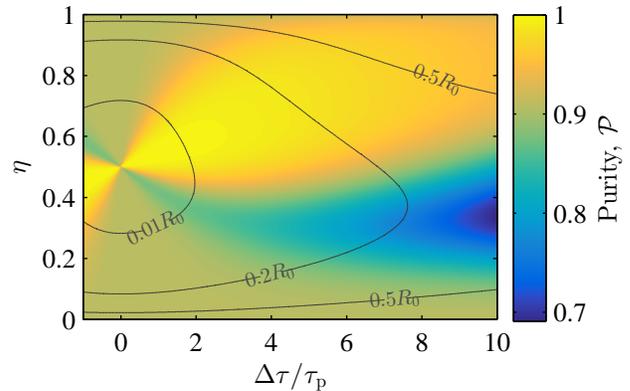}
\caption{Purity $\mathcal{P}$, as a function of the dual-pulse splitting ratio $\eta$ and temporal separation $\Delta \tau$, in the case of $\tau_p \Omega =1/5$.  }
\label{fig3}
\end{center}
\end{figure}

Figure \ref{fig:2} demonstrates the advantage of our dual-pulse scheme for a particular parameter set ($\eta, \Delta \tau$) from Eq.~\eqref{eq:dualpulse}. In the following, we investigate how the purity depends on these two parameters. Figure \ref{fig3} shows $\mathcal{P}(\eta, \Delta \tau)$ for the case of $\tau_\p \Omega = 1/5$. We note that it is qualitatively similar for other values of $\tau_\p \Omega $, and that the purity is mirror symmetric in the sense that $\mathcal{P}(\eta, \Delta \tau) =  \mathcal{P}(1-\eta, -\Delta \tau)$ as is evident from Eq.~\eqref{eq:dualpulse}. Promisingly, a large subset of the $(\eta, \Delta \tau)$-parameter space results in $\mathcal{P} > 0.92$, which is an improvement in comparison to the single-pump case recovered in the extremes $\eta = 0$ or $\eta = 1$. Furthermore, Fig.~\ref{fig3} reveals a large area in the $(\eta, \Delta \tau)$-parameter space for which $\mathcal{P}>0.99$. Such robustness is vital to the experimental feasibility of the scheme.

\begin{figure}[t]
\begin{center}
\includegraphics[scale=1]{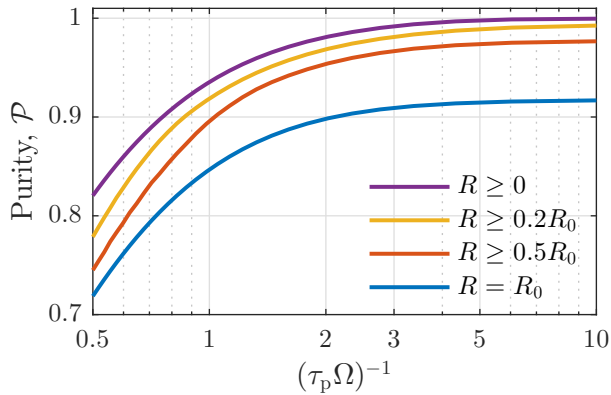}
\caption{Purity $\mathcal{P}$, as a function of the normalized spectral pump width $(\tau_\p \Omega)^{-1}$ with and without constraints on the relative pair-generation probability $R/R_0$, where $R_0$ is the generation probability in the single-pump case.}
\label{fig4}
\end{center}
\end{figure}

The increase in purity allowed by our dual-pulse scheme, comes at the cost of having less pump power in the resonator, and hence a lower generation rate $R$. This dependence is illustrated by the contours in Fig.~\ref{fig3}, which represent different generation rates $R$ relative to the single-pump case, for which we denote the rate $R_0$. These contours show how the generation rate increases as we move away from the point $(\eta = 1/2, \Delta  \tau = 0)$, for which $R=0$ [see Eq.~\eqref{eq:dualpulse}]. In practice, given an MRR with quality factor $Q$ and a laser-pulse duration $\tau_\p$, the purity should be optimized under some constraint on the desired generation rate $R$. Figure \ref{fig4} shows the maximal purity as a function of $\tau_\p$ under different generation-rate constraints. The bottom (blue) curve shows the single-pulse case, which saturates at 0.92 for pump pulses that are spectrally much broader than the resonance linewidth. For each value of $\tau_\p$, we denote the corresponding generation rate $R_0$, permitting us to evaluate the penalty in generation rate when using the dual-pulse scheme. The remaining curves were obtained by numerically maximizing $\mathcal{P}(\eta, \Delta \tau)$ (see Fig.~\ref{fig3}) subject to the constraints $R/R_0 \geq 0.5$ (red), $ R/R_0 \geq 0.2$ (yellow), and $R \geq 0$ (purple, unconstrained). Remarkably, a purity in excess of 0.99 can be achieved with $R > 0.2 R_0$; a penalty in generation rate that, due to the quadratic scaling between pair-generation rate and pulse energy, can be compensated by an increase of a factor $2.2$ in pulse energy.



\textit{Experimental considerations}.---As seen from Fig.~\ref{fig3}, our scheme is not critically dependent on the choice of the parameters $\eta$ and $\Delta \tau$. In fact, a large region in the parameter space enables the generation of heralded single photons with purities far exceeding those attainable with a single pump pulse. However, from an optimization point-of-view one wishes to minimize spectral correlation given some constraint on the available pump power (or constraint on photon-pair generation rate). To this end, high-resolution spectral-correlation characterization can be performed by means of classical stimulated four-wave mixing enabling sub-hour measurement of the joint spectral intensity \cite{liscidini2013}. Even though the spectral features of MRRs demand highly spectrally resolving equipment, this technique has recently been demonstrated for a silicon MRR photon-pair source \cite{grassani2016}.

\begin{figure}[h]
\begin{center}
\includegraphics[scale=1]{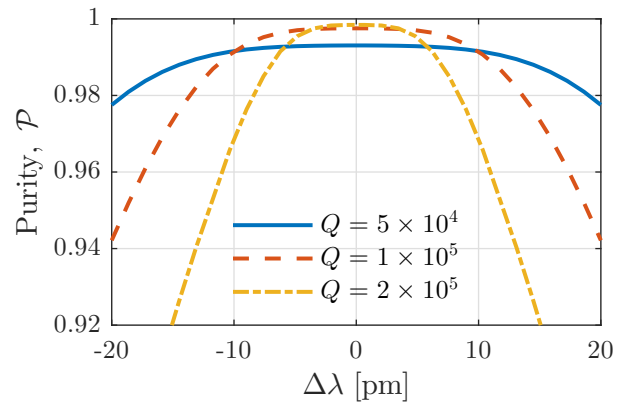}
\caption{Purity $\mathcal{P}$, as a function of ring-resonance wavelength shift $\Delta \lambda$. Case of $\eta =0.6$, $\tau_\p = 10~\mathrm{ps}$, and $\Delta \tau \Omega = 0.3$, and a pump wavelength centered at 1550 nm. }
\label{fig5}
\end{center}
\end{figure}

An immediate advantage of the dual-pulse scheme is that it can readily be implemented in already existing MRR-based photon-pair sources. Such implementation requires a preparation setup comprised of an unbalanced Mach-Zehnder interferometer, in which the splitting ratio of the first beam splitter is $\eta$, the path difference is $\Delta \tau$, and the combining beam splitter is 50/50. Such an interferometer, which potentially could be integrated directly on-chip, would directly generate the spectrum given by Eq.~\eqref{eq:dualpulse} in one of its output ports. This, however, assumes perfect phase-stability of the two interferometric arms, something which in reality can not be achieved. More generally, we therefore ought to consider a relative phase of $\phi \neq \pi$ between the two terms in Eq.~\eqref{eq:dualpulse}, which would have the effect of changing the position of the pulse spectral dip relative to the resonance. Moreover, the resonance might itself shift, which likewise induces asymmetry between the pulse spectrum and the resonance spectrum, and therefore has the same qualitative effect. Figure \ref{fig5} shows how the purity depends on the resonance shift $\Delta \lambda$ for different resonator quality factors. In agreement with intuition, a larger quality factor entails the need for improved phase-control over the ring. 

\textit{Conclusion}.---We have presented a method for generating spectrally unentangled photon pairs using spontaneous four-wave mixing in standard microring resonators. The developed method requires an input pump spectrum attainable by using two pulses which are slightly separated in time with a $\pi$-relative phase shift. Such a pump configuration enables a broadened in-resonator pump spectrum, which is necessary for completely eliminating signal-idler spectral correlations. We further show that the proposed scheme is highly robust with respect to realistic experimental parameter uncertainties, making it a promising candidate for future on-chip implementation of high-purity heralded single photons.






\begin{acknowledgments}
The authors were supported by: DFF Sapere Aude Adv.~Grant NANO-SPECs; DFF (Grant No.~4184-00433); and the DNRF Research Centre of Excellence, SPOC  (ref.~DNRF123).
\end{acknowledgments}



%




\bibliography{biblo}

\end{document}